\documentclass[12pt]{iopart}
\usepackage{graphicx}
\begin{document}
\title[]{Current status of Japanese detectors}
\author{
Daisuke Tatsumi$^1$, Ryutaro Takahashi$^1$, Koji Arai$^1$,
Noriyasu Nakagawa$^2$, Kazuhiro Agatsuma$^2$, 
Toshitaka Yamazaki$^1$, Mitsuhiro Fukushima$^1$, Masa-Katsu Fujimoto$^1$,
\
Akiteru Takamori$^3$, 
\
Alessandro Bertolini$^4$, 
\
Virginio Sannibale$^5$, Riccardo DeSalvo$^5$, 
\
Szabolcs M\'{a}rka$^6$
\
Masaki Ando$^7$, Kimio Tsubono$^7$ 
\
Tomomi Akutsu$^2$, Kazuhiro Yamamoto$^2$, Hideki Ishitsuka$^2$, 
Takashi Uchiyama$^2$, Shinji Miyoki$^2$, Masatake Ohashi$^2$, 
Kazuaki Kuroda$^2$,
\
Norichika Awaya$^8$, Nobuyuki Kanda$^8$
\
Akito Araya$^3$,
\
Souichi Telada$^9$, 
\
Takayuki Tomaru$^{10}$, Tomiyoshi Haruyama$^{10}$, Akira Yamamoto$^{10}$, 
Nobuaki Sato$^{10}$, Toshitaka Suzuki$^{10}$ and Takakazu Shintomi$^{11}$
}
\address{
$^1$ National Astronomical Observatory of Japan, Mitaka, Tokyo 181-8588, Japan \\
$^2$ Institute for Cosmic Ray Research, University of Tokyo, Kashiwa, Chiba 277-8582, Japan \\
$^3$ Earthquake Research Institute, The University of Tokyo, Bunkyo-ku, Tokyo 113-0032, Japan \\
$^4$ Universit`a di Pisa, Pisa, Italy \\
$^5$ California Institute of Technology, MS 18-34, Pasadena CA 91125, USA \\
$^6$ Columbia University, 2960 Broadway, New York, USA \\
$^7$ Department of Physics, The University of Tokyo, Bunkyo-ku, Tokyo 113-0033, Japan \\
$^8$ Department of Physics, Graduate School of Science, Osaka City University, Sumiyoshi-ku, Osaka, JAPAN \\
$^9$ National Institute for Advanced Industrial Science and Technology, Umezono, Tsukuba, Ibaraki, Japan \\
$^{10}$ High Energy Accelerator Research Organization (KEK), Tsukuba, Ibaraki 305-0801, Japan \\
$^{11}$  Advanced Research Institute for the Sciences and Humanities, Nihon University, Chiyoda-ku, Tokyo, Japan
}
\eads{daisuke.tatsumi@nao.ac.jp}

\begin{abstract}
Current status of TAMA and CLIO detectors in Japan is reported in this article.
These two interferometric gravitational-wave detectors are being developed for the large cryogenic gravitational wave telescope (LCGT) which is a future plan for detecting gravitational wave signals at least once per year.

TAMA300 is being upgraded to improve the sensitivity in low frequency region after the last observation experiment in 2004. To reduce the seismic noises, we are installing new seismic isolation system, which is called TAMA Seismic Attenuation System, for the four test masses. We confirmed stable mass locks of a cavity and improvements of length and angular fluctuations by using two SASs. We are currently optimizing the performance of the third and fourth SASs. We continue TAMA300 operation and R\&D studies for LCGT. Next data taking in the summer of 2007 is planned.

CLIO is a 100-m baseline length prototype detector for LCGT to investigate interferometer performance in cryogenic condition. The key features of CLIO are that it locates Kamioka underground site for low seismic noise level, and adopts cryogenic Sapphire mirrors for low thermal noise level.
The first operation of the cryogenic interferometer was successfully demonstrated in February of 2006. Current sensitivity at room temperature is close to the target sensitivity within a factor of 4. Several observation experiments at room temperature have been done. Once the displacement noise reaches at thermal noise level of room temperature, its improvement by cooling test mass mirrors should be demonstrated.
\end{abstract}

Current status of TAMA and CLIO detectors in Japan is reported in this article.
These two interferometric gravitational wave detectors are being developed for LCGT which is a future plan for detecting gravitational wave signals at least once per year.

\section{TAMA}
The TAMA project held observation experiments for detecting gravitational wave (GW) signals since 1999. By the beginning of 2004, 3000 hours of data in total was accumulated through the nine observation experiments.

After the last observation experiment in 2004, TAMA detector is being upgraded to reduce the low frequency noises.
To reduce the seismic noise, new isolation system called TAMA Seismic Attenuation System (SAS) \cite{tama1, tama2} is being installed. To isolate horizontal motions, an inverted pendulum is employed. For vertical motion, double stage Monolithic Geometric Anti-Spring (MGAS) filters are used. The test mass mirror is suspended by a double pendulum as a final stage of the system. 
The SAS installation was started in September, 2005.

\begin{figure}[bht*]
   \begin{center}
   \includegraphics[scale=0.95,angle=0.0]{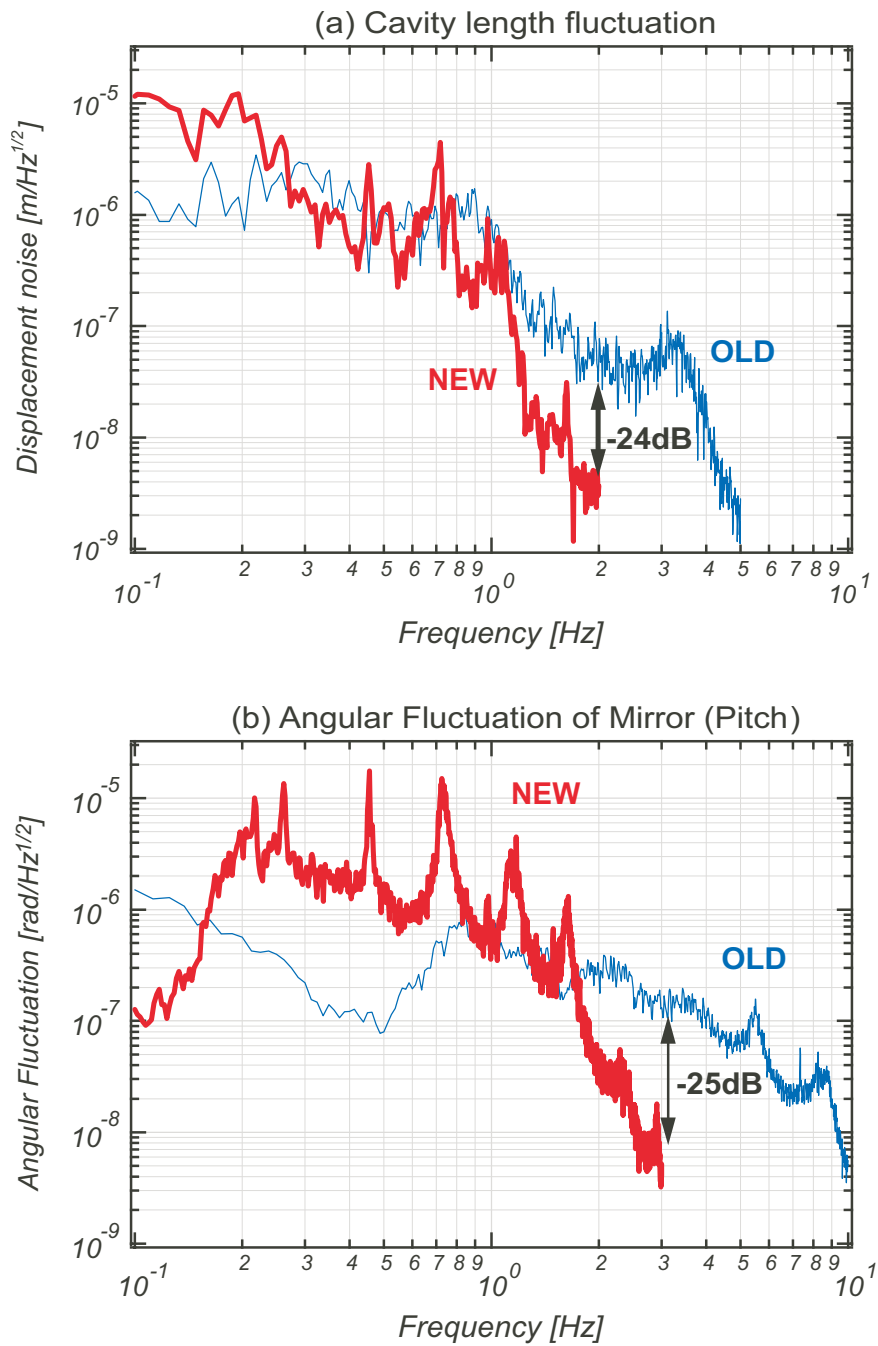}
   \caption{
   Measured the mirror fluctuations of TAMA: (a) a cavity length fluctuation and (b) an angular fluctuation in pitch of the mirror which is suspended by SAS.
    The improvement of about 24dB was observed above 2 and 3 Hz for the length and pitch motion, respectively.
   } \label{tama-fig1}
   \end{center}
\end{figure}
In the summer of 2006, a 300-m Fabry-Perot cavity formed by the mirrors suspended by SASs was operated. Since the other arm was still formed by the old suspension system, we could directly compare the performance of the vibration isolation systems. In this locked Fabry-Perot configuration, the old suspension arm was controlled by the laser frequency, and the SAS arm was controlled by the coil-magnet actuators located at test mass mirror. Figure \ref{tama-fig1} shows the comparison of cavity length and angular fluctuations with SASs and the old suspension systems. The improvement of about 24dB was observed above 2 and 3 Hz for the length and pitch motion, respectively.

The actuator response of the test mass for SAS was reduced by a factor of 3 so as to reduce the actuator associated noises. Therefore, ability of the lock acquisition and drift level of the interferometer with SASs should have been confirmed. 
To ensure the ability of the lock acquisition, a digital servo system was employed for the test mass length control. 
This system is driven by a DSP processor with sampling frequency of 200 kHz, and the unity gain frequency of about 800Hz was achieved. This system utilizes two techniques for the error signal operation. The first is gating of the error signal; the error signal is forced to be zero when the transmitted light level of the cavity is below a preset threshold. This cleans up undesirable signals caused by resonances of modulation sidebands and higher order spatial modes of the beam. The other technique is normalization of the error signal with regard to the transmitted light. This widened linear range of the error signal by a factor of 3.
The locked Fabry-Perot interferometer was locked with help of the digital servo system, and operated continuously for the 6.5 hours. 

Since the detected length fluctuation is the addition of that of the SAS arm and the old suspension arm, this means that the actuator range was enough to compensate the drift motion of the SASs. This indicates that TAMA300 would be operable even with both two arms are formed with SASs.

Actual improvements of length and angular fluctuations at 100 Hz region should be confirmed by the locked Fabry-Perot configuration formed by four SASs. 
Now all of the four test mass mirrors are suspended by SASs. 
We are currently optimizing the performance of the third and fourth SASs. 
We continue R\&D studies of SAS and TAMA300 operations. Next data taking in the summer of 2007 is planned.

\section{CLIO}
Cryogenic Laser Interferometer Observatory (CLIO) is located in Kamioka mine of Japan as an underground site. Purpose of CLIO project is technical demonstration of key features of LCGT. LCGT is planed to locate at Kamioka underground site for low seismic noise level, to adopt cryogenic Sapphire mirrors for low thermal noise level and to have arms of 3km long. Except for the arm length, CLIO has same features of LCGT. Therefore, the detector can demonstrate them as a prototype of LCGT.

Construction of CLIO has been started in 2002.
All of vacuum pipes, cryostats and cryocoolers were installed by the June, 2005 \cite{clio1}.
After the cooling tests of the system, the first operation of the cryogenic interferometer has been demonstrated on 18 February, 2006. 
The details of the interferometer, cryogenic systems and seismic environments have been published in Ref. \cite{clio2, clio3, clio4, clio5}.
Input power to each cavity was 200m W. The measured finesse and reflectance of the cavity were about 3100 and 0.01, respectively. The accumulated power inside the cavity was estimated to be 200 W.
During the lock of 50 minutes, the mirrors kept its temperature around 20K.

\begin{figure}[htb*]
   \begin{center}
   \includegraphics[scale=0.59,angle=0.0]{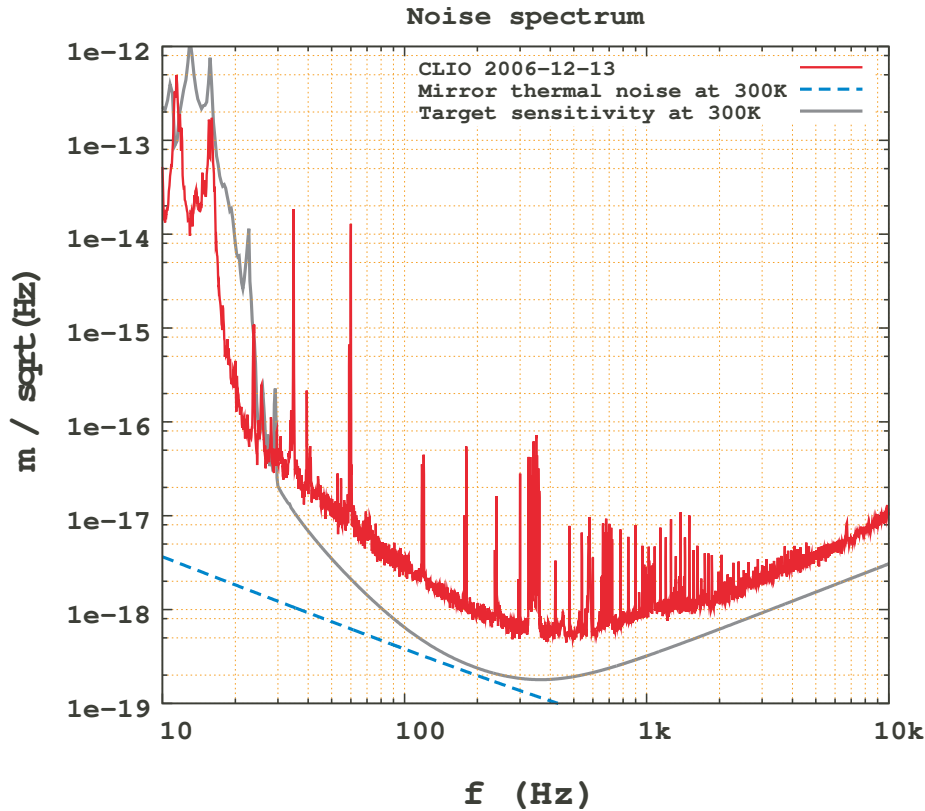}
   \caption{
   Noise spectrum of CLIO detector on 13 December, 2006.
   At the time, the mirror temperature was about 300K.
   At all of frequency regions, the differences from the target sensitivity at room temperature are about a factor of 4.
   } \label{clio-fig2}
   \end{center}
\end{figure}

After the cryogenic operation was established, noise investigations at room temperature are in progress. Figure \ref{clio-fig2} shows the best noise spectrum on 13 December 2006. At all of frequency regions, the differences from the target sensitivity at room temperature are about a factor of 4.

\begin{figure}[htb*]
   \begin{center}
   \includegraphics[scale=0.75,angle=0.0]{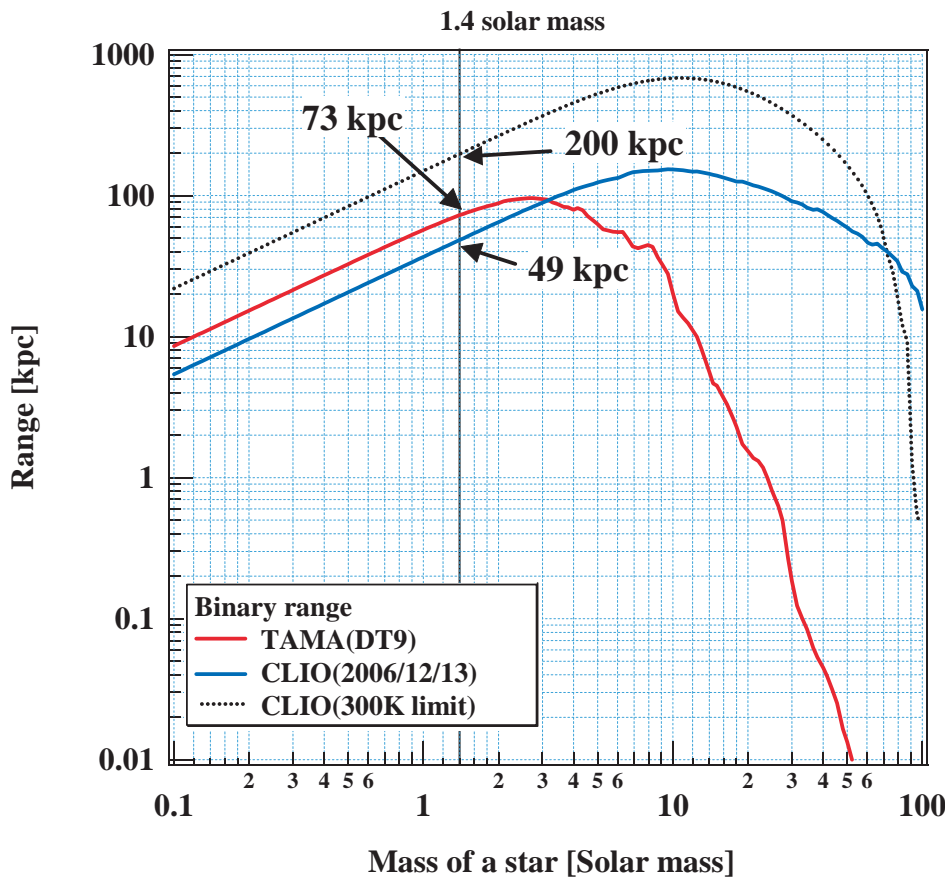}
   \caption{
   The observable ranges with SNR=10 for inspiral GW signals estimated by the noise spectra. For neutron star binaries, CLIO and TAMA can observe the event within 49k pc and 73k pc, respectively. We can say that the two detectors have almost same sensitivity. At over 10 solar mass regions, CLIO keeps good sensitivities due to its low seismic noises. It is the greatest benefit of underground site.
   } \label{clio-fig3}
   \end{center}
\end{figure}

The observable ranges with SNR=10 for inspiral GW signals can be estimated by the noise spectra as shown in Fig. \ref{clio-fig3}. For neutron star binaries, CLIO and TAMA can observe the event within 49k pc and 73k pc, respectively. We can say that the two detectors have almost same sensitivity. At over 10 solar mass regions, CLIO keeps good sensitivities due to its low seismic noises. It is the greatest benefit of underground site. 

Several observation experiments at room temperature have been done.
Once the displacement noise reaches at thermal noise level, its improvement by cooling should be demonstrated.

\ack
TAMA and CLIO projects are supported in part by a Grant-in-Aid for Scientific Research on Priority Areas (415) of the Ministry of Education, Culture, Sports, Science and Technology and Grant-in-Aid for Scientific Research (A), No.18204021, 2006 of Japan Society for the Promotion of Science.

Developments of TAMA SAS were supported by the Advanced Technology Center of the National Astronomical Observatory of Japan and the U.S. National Science Foundation under Cooperative Agreement No. PHY-0107417.

\end{document}